# Characterization of passivity in Mueller matrices

**IGNACIO SAN JOSÉ[1], JOSÉ J. GIL[2*]**

[1] *Instituto Aragonés de Estadística, Gobierno de Aragón, Bernardino Ramazzini 5, 50015 Zaragoza, Spain*
[2] *Department of Applied Physics, University of Zaragoza, Pedro Cerbuna 12, 50009 Zaragoza, Spain*

*Corresponding author: ppgil@unizar.es

**Except for very particular and artificial experimental configurations, linear transformations of the state of polarization of an electromagnetic wave result in a reduction of the intensity of the exiting wave with respect to the incoming one. This natural passive behavior imposes certain mathematical restrictions on the corresponding Mueller matrices associated to the said transformations. Although the general conditions for passivity in Mueller matrices were presented in a previous paper [J. J. Gil, J. Opt. Soc. Am. A 17, 328-334 (2000)], the demonstration was incomplete. In this paper, the set of two necessary and sufficient conditions for a Mueller matrix to represent a passive medium are determined and demonstrated on the basis of its arbitrary decomposition as a convex combination of nondepolarizing and passive pure Mueller matrices. The procedure followed to solve the problem provides also an appropriate framework to identify the Mueller matrix that, among the family of proportional passive Mueller matrices, exhibits the maximal physically achievable intensity transmittance. Beyond the theoretical interest on the rigorous characterization of passivity, the results obtained, when applied to absolute Mueller polarimetry, also provide a criterion to discard those experimentally measured Mueller matrices that do not satisfy the passivity criterion.**

## 1. INTRODUCTION

Polarimetry constitutes today a very dynamic area in science and engineering that involves powerful measurement techniques widely exploited for the study and analysis of great variety of material samples. Consequently, the mathematical characterization of the polarimetric properties of material media has a capital interest because it provides tools for the analysis and interpretation of experimental measurements. The appropriate framework for the mathematical representation of linear polarization interactions is given by the Stokes-Mueller formalism. Mueller matrices are 4×4 real matrices that perform the linear transformation from the Stokes parameters of the incoming state of polarization to the outgoing one. The physical nature of such linear interactions imposes certain restrictions that are reflected in the fact that the set of Mueller matrices is constituted by a specific subset of real 4×4 matrices.

The Mueller-Stokes transformations are determined by an ensemble average (a convex sum) of basic pure transformations (*ensemble criterion*) [1,2], each one characterized by a well-defined Mueller-Jones matrix (also called *pure* or *nondepolarizing* Mueller matrix). This feature leads to the *covariance criterion* that was mathematically formulated by Cloude [3] and, independently, by Arnal [4], through the nonnegativity of the four eigenvalues of the covariance matrix **H** associated with a given Mueller matrix **M** (thus providing four *covariance inequalities* to be satisfied by the elements of **M**).

A complementary criterion refers to passivity and implies that the action of the medium does not amplify the intensity of the electromagnetic wave interacting with it. More specifically, the assumption of the ensemble criterion entails the necessity that a passive Mueller matrix is susceptible to be expressed as a convex combination of pure and passive Mueller matrices. This fact is what should be mathematically formulated in order to obtain the passivity conditions to be satisfied by **M**.

Leaving aside certain artificial arrangements where the medium involves intensity amplifiers [5], both natural and man-made objects do not amplify the intensity of light, but generally reduce it to some extent. As limiting situations, transparent systems correspond to the ideal case of media that preserve the intensity, while opaque systems produce zero output intensity (so that they are polarimetrically represented by the zero Mueller matrix). Polarimetric techniques usually deal with the measurement and characterization of the polarization properties of a great variety of material targets in science, industry, medicine, remote sensing, etc., where the samples are inherently passive. Thus, passivity is a physical condition that must be taken into account in the mathematical characterization of the polarimetric properties of material media.

The passivity criterion has been dealt with by several authors from long time ago, providing relevant results. Nevertheless, although the *forward* and *reverse* passivity conditions for general Mueller matrices (either nondepolarizing or depolarizing) were established in a previous paper [6], the demonstration of the sufficiency of such necessary conditions was not performed in a complete way. Furthermore, the inspection of the type-II canonical form of a Mueller matrix [7], made us think that the above-mentioned passivity conditions are not sufficient [8]. The origin of the said controversy came from the formulation of the *arbitrary decomposition* [9-11] of a Mueller matrix with the unnecessary exigency that all the pure components have the same value for the mean intensity coefficient. In what follows, we will show that such constraint is not necessary and, by means of a proper demonstration, we will found that the conditions stated in [6] are correct and apply to any kind of passive system, thus determining definitively the general characterization of passive Mueller matrices.

The approach to the problem is based on revisiting the well-known conditions for a Mueller matrix to represent a passive medium (including the simple demonstration that they are necessary) and then demonstrate that such conditions are also sufficient.

In order to formulate the problem, it is worth to bring up the partitioned block expression of a Mueller matrix [12], which will be used for both pure and general (depolarizing) Mueller matrices.

$$\mathbf{M} = m_{00}\hat{\mathbf{M}}, \quad \hat{\mathbf{M}} \equiv \begin{pmatrix} 1 & \mathbf{D}^T \\ \mathbf{P} & \mathbf{m} \end{pmatrix},$$

$$\mathbf{m} \equiv \frac{1}{m_{00}}\begin{pmatrix} m_{11} & m_{12} & m_{13} \\ m_{21} & m_{22} & m_{23} \\ m_{31} & m_{32} & m_{33} \end{pmatrix}, \quad (1)$$

$$\mathbf{D} \equiv \frac{(m_{01}, m_{02}, m_{03})^T}{m_{00}}, \quad \mathbf{P} \equiv \frac{(m_{10}, m_{20}, m_{30})^T}{m_{00}},$$

where the superscript $T$ indicates transpose, $m_{00}$ is the *mean intensity coefficient* (MIC) (i.e. the *transmittance* or *gain* [13-17] of **M** for input unpolarized light), and **D** and **P** are the respective diattenuation and polarizance vectors of **M**. The magnitudes of these vectors are the diattenuation $D \equiv |\mathbf{D}|$ and the polarizance $P \equiv |\mathbf{P}|$. An overall combined measure of diattenuation-polarizance is given by the degree of polarizance $P_P$ defined as [18]

$$P_P \equiv \sqrt{(D^2 + P^2)/2}. \quad (2)$$

Given the peculiar mathematical structure of a pure Mueller matrix $\mathbf{M}_J$, its transposed matrix $\mathbf{M}_J^T$ is also a pure Mueller matrix [19,20]. Thus, by virtue of the arbitrary decomposition of a depolarizing Mueller matrix **M** into a convex sum of pure Mueller matrices, it follows that $\mathbf{M}^T$ is necessarily a Mueller matrix.

Let us now consider the pair of Stokes vectors $(1, \mathbf{D}^T)^T$ and $(1, \mathbf{P}^T)^T$ as respective input vectors for matrices **M** and $\mathbf{M}^T$, so that the intensities $s'_0$ and $s''_0$ of the respective output Stokes vectors $\mathbf{M}(1, \mathbf{D}^T)^T$ and $\mathbf{M}^T(1, \mathbf{P}^T)^T$ are given by $s'_0 = m_{00}(1+D)$ and $s''_0 = m_{00}(1+P)$. Since the input intensities are 1, it follows that the passivity of **M** (hence of $\mathbf{M}_J^T$ and vice versa) entails the conditions [6]

$$m_{00}(1+D) \leq 1, \quad m_{00}(1+P) \leq 1, \quad (3)$$

which therefore are necessary for **M** to be passive.

In order to get a constructive demonstration of the fact that the passivity conditions (3) are also sufficient for **M** to be passive we will organize this paper in the following way. In Sec. 2, the passivity condition for pure Mueller matrices is retrieved; then, in Sec. 3, the generalized arbitrary decomposition of a Mueller matrix **M** into sets of pure Mueller matrices is formulated; then, to simplify further calculations, it is defined in Sec. 4 the tridiagonal form of **M** as well as the canonical passive form $\tilde{\mathbf{M}}$ of **M**; Sec. 5 is devoted to show that the limiting situation for passivity occurs when the pure arbitrary components of $\hat{\mathbf{M}}$ have all respective diattenuation or polarizance vectors parallel to those of $\hat{\mathbf{M}}$; the general form of a pure Mueller matrix satisfying such vector condition is obtained in Sec. 6; then the desired general demonstration is performed in Sec. 7 in terms of the rank of the coherency matrix **C** associated with $\tilde{\mathbf{M}}$. Note that this Section 1 is merely introductory and that the notions involved will become clear as the consecutive sections are developed.

## 2. PASSIVITY CONDITION FOR PURE MUELLER MATRICES

Let us first recall that any macroscopic interaction of light with matter always can be considered as the result of a composition of a number of basic molecular interactions, each one, taken isolated, being necessarily nondepolarizing (that is, never producing a reduction of the degree of polarization of incoming fully polarized light). For each nondepolarizing element, its polarimetric properties are fully determined either by means of the corresponding Jones matrix **T**, either through the associated Mueller matrix $\mathbf{M(T)}$. While **T** is a 2x2 complex matrix that transforms the input polarization matrix **Φ** (representing the state of polarization of the incoming light), into the output polarization matrix $\mathbf{\Phi'} = \mathbf{T\Phi T}^\dagger$ (associated with the outgoing light), where the dagger stands for complex conjugate, its corresponding pure Mueller matrix $\mathbf{M(T)}$ is a 4×4 real matrix of the form

$$\mathbf{M(T)} \equiv \mathcal{L}(\mathbf{T} \otimes \mathbf{T}^*)\mathcal{L}^{-1},$$

$$\mathcal{L} \equiv \frac{1}{\sqrt{2}}\begin{pmatrix} 1 & 0 & 0 & 1 \\ 1 & 0 & 0 & -1 \\ 0 & 1 & 1 & 0 \\ 0 & i & -i & 0 \end{pmatrix}. \quad (4)$$

where $\otimes$ indicates Kronecker product.

Let us first consider the passivity criterion for Jones matrices, which will determine the corresponding criterion for pure Mueller matrices. Any 2×2 complex matrix can be considered a Jones matrix, except with respect to passivity. The condition for **T** to represent a passive nondepolarizing medium arises from the physical restriction that the ratio between the intensities of the emerging and incident beams must be less than 1, which leads to the following necessary and sufficient passivity condition [13]

$$p_1^2 \equiv \frac{1}{2}\text{tr}(\mathbf{T}^\dagger\mathbf{T})\left\{1 + \sqrt{1 - \frac{4\det(\mathbf{T}^\dagger\mathbf{T})}{[\text{tr}(\mathbf{T}^\dagger\mathbf{T})]^2}}\right\} \leq 1. \quad (5)$$

In fact, the above quantity is not other than the square of the largest singular value $p_1$ of **T**. The singular value decomposition of **T** can be expressed as [21]

$$\mathbf{T} = \mathbf{T}_{R2}\,\text{diag}(p_1, p_2)\,\mathbf{T}_{R1}, \quad (6)$$

where $\mathbf{T}_{R1}$ and $\mathbf{T}_{R2}$ are unitary matrices and $\mathbf{T}_{DL0} \equiv \text{diag}(p_1, p_2)$ is a diagonal matrix whose diagonal elements are the real nonnegative singular values $p_1$ and $p_2$.

In the case of pure Mueller matrices, due to their peculiar structure, the equality $P = D$ is always satisfied [19], so that $p_1^2 = m_{00}(1+D) = m_{00}(1+P)$ and the passivity condition (5) adopts the simple form

$$m_{00}(1+D) \leq 1, \quad (P = D). \quad (7)$$

## 3 ARBITRARY DECOMPOSITION OF A MUELLER MATRIX

In order to characterize the passivity of depolarizing Mueller matrices it is necessary to revisit some important concepts concerning their structure.

From the ensemble criterion it follows that, given a Mueller matrix **M**, its associated *covariance matrix* **H** is defined as [3,4]

$$\mathbf{H(M)} = \frac{1}{4}\sum_{i,j=0}^{3} m_{ij}(\boldsymbol{\sigma}_i \otimes \boldsymbol{\sigma}_j), \quad (8)$$

where $\boldsymbol{\sigma}_i$ are the Pauli matrices (taken in the order commonly used in polarization optics)

$$\boldsymbol{\sigma}_0 = \begin{pmatrix} 1 & 0 \\ 0 & 1 \end{pmatrix},\ \boldsymbol{\sigma}_1 = \begin{pmatrix} 1 & 0 \\ 0 & -1 \end{pmatrix},\ \boldsymbol{\sigma}_2 = \begin{pmatrix} 0 & 1 \\ 1 & 0 \end{pmatrix},\ \boldsymbol{\sigma}_3 = \begin{pmatrix} 0 & -i \\ i & 0 \end{pmatrix}. \quad (9)$$

**H** is positive-semidefinite, that is, the four eigenvalues of **H** are nonnegative. Conversely, the elements of **M** can be expressed as follows as functions of **H**

$$m_{ij} = \mathrm{tr}\left[ (\boldsymbol{\sigma}_i \otimes \boldsymbol{\sigma}_j) \mathbf{H} \right]. \quad (10)$$

It is worth to observe that any unitary similarity transformation of **H**, $\mathbf{VHV}^\dagger$ with $\mathbf{V}^\dagger = \mathbf{V}^{-1}$, constitutes an alternative positive semidefinite Hermitian matrix that also contains all the polarimetric information of the medium, and therefore can be used as its representative. Among these possible covariance matrices, for certain calculations it is sometimes useful to consider the so-called *coherency matrix* **C** [3], linked to **H** through the similarity transformation

$$\mathbf{C}(\mathbf{M}) = \mathcal{L}\left[ \mathbf{H}(\mathbf{M}) \right] \mathcal{L}^{-1}. \quad (11)$$

Note that $\mathrm{rank}(\mathbf{C}) = \mathrm{rank}(\mathbf{H}) \equiv r$, $r$ being the minimum number of pure incoherent components of **M** [10,11]. The explicit expressions for $\mathbf{H}(\mathbf{M})$, $\mathbf{M}(\mathbf{H})$, $\mathbf{C}(\mathbf{M})$ and $\mathbf{M}(\mathbf{C})$ can be found in [21,22].

The reason for the formulation of the problem in terms of coherency matrices comes from the fact that their peculiar structure (diagonal Mueller matrices have associated diagonal coherency matrices), makes them simpler certain calculations to be carried out for the demonstration that conditions (3) are sufficient for a Mueller matrix to be passive.

Since **C** is a positive semidefinite Hermitian matrix [3], it can be diagonalized as

$$\mathbf{C} = \mathbf{U}\,\mathrm{diag}(\lambda_0, \lambda_1, \lambda_2, \lambda_3)\,\mathbf{U}^\dagger, \quad (12)$$

where $\lambda_i$ are the four non-negative eigenvalues of **C**, taken in decreasing order ($0 \le \lambda_3 \le \lambda_2 \le \lambda_1 \le \lambda_0$). The columns $\mathbf{u}_i$ ($i = 0,1,2,3$) of the 4×4 unitary matrix **U** are the respective unit, mutually orthogonal, eigenvectors.

Therefore, **C** can be expressed as the following convex linear combination of four rank-1 coherency matrices that represent respective pure systems

$$\mathbf{C} = \sum_{i=1}^{r} \frac{\lambda_i}{m_{00}} \mathbf{C}_{Ji},\ \ \mathbf{C}_{Ji} \equiv m_{00}\left( \mathbf{u}_i \otimes \mathbf{u}_i^\dagger \right),\ \ m_{00} = \mathrm{tr}\,\mathbf{C}. \quad (13)$$

This (*Cloude decomposition* [3], or *spectral decomposition*) can be written in terms of the corresponding Mueller matrices by means of the following convex sum

$$\mathbf{M} = \sum_{i=1}^{r} \frac{\lambda_i}{m_{00}} \mathbf{M}_{Ji},\ \ (\mathbf{M}_{Ji})_{00} = m_{00} = \mathrm{tr}\,\mathbf{C}, \quad (14)$$

where all pure Mueller matrices $\mathbf{M}_{Ji}$ have equal MIC, equal to $m_{00}$. Hereafter, when appropriate, pure Mueller matrices and pure coherency matrices will be denoted as $\mathbf{M}_J$ and $\mathbf{C}_J$ respectively.

While the components of the spectral decomposition are defined from the respective eigenvectors $\mathbf{u}_i$ of **C**, any Mueller matrix also admit the so-called *arbitrary decomposition* [10,11]

$$\mathbf{M} = \sum_{i=1}^{r} p_i \mathbf{M}_{Ji},\ \ (\mathbf{M}_{Ji})_{qr} = m_{00}\,\mathrm{tr}\left[ (\boldsymbol{\sigma}_q \otimes \boldsymbol{\sigma}_r)(\hat{\mathbf{w}}_i \otimes \hat{\mathbf{w}}_i^\dagger) \right],$$

$$p_i = \frac{1}{m_{00}\sum_{j=1}^{r}\frac{1}{\lambda_j}\left|\left(\mathbf{U}^\dagger \hat{\mathbf{w}}_i\right)_j\right|^2},\ \ \sum_{i=1}^{r} p_i = 1, \quad (15)$$

where $\hat{\mathbf{w}}_i$ ($i = 1,...,r$) is a set of $r$ independent unit vectors belonging to the image subspace of **C** [denoted as $\mathrm{range}(\mathbf{C})$] [11]. Note that when $\hat{\mathbf{w}}_i = \hat{\mathbf{u}}_i$ ($\hat{\mathbf{u}}_i$ being the unit eigenvectors of **C** with nonzero eigenvalue), then the arbitrary decomposition adopts the particular form of the spectral decomposition. Decompositions (14) and (15) have been formulated with all pure components having MIC equal to $m_{00}$. Nevertheless, they can be generalized as follows to the case where the MIC $m_{00i}$ of the said pure components are different [23]

$$\mathbf{M} = \sum_{i=1}^{r} k_i \mathbf{M}_{Ji},$$

$$k_i = \frac{m_{00}}{m_{00i}} p_i = \frac{1}{m_{00i}\sum_{j=1}^{r}\frac{1}{\lambda_j}\left|\left(\mathbf{U}^\dagger \hat{\mathbf{w}}_i\right)_j\right|^2}, \quad (16)$$

$$\left( \sum_{i=1}^{r} k_i = \sum_{i=1}^{r} p_i = 1 \right).$$

Some examples of parallel compositions of pure Mueller matrices having different respective MIC can be found in [23,24].

## 4. PASSIVE FORM OF A MUELLER MATRIX

It is frequent that Mueller polarimetry setups provide the Mueller matrix **M** of the sample up to a positive scale factor (*relative* Mueller polarimetry). Nevertheless, the *absolute* (or *complete*) measurement of the sixteen elements of **M**, thus including its corresponding MIC $m_{00}$, is interesting in general because $m_{00}$, together with other elements of **M**, holds physical information on the polarization-dependent transmittance of the medium represented by **M**. For instance, when, up to the tolerance-precision of the polarimeter, the measured **M** corresponding to a passive medium does not satisfy the necessary passivity conditions (3), this indicates that the polarimeter is not working properly, and that such particular measured **M** should be discarded because of the lack of compatibility between theory and experiment. Furthermore, it is common that the experimentalist uses some hypothesis about one or more parallel components of **M** [25,26], so that the passivity criterion may become important in order to check the physical realizability of the decomposition or polarimetric subtraction performed [11,27]. In other words, in addition to the Cloude's criterion [3], passivity provides a way to admit or discard the physical realizability of a measured **M** as well as its possible parallel decompositions. Indeed, the interest of considering absolute polarimetry as well as the physical and mathematical constraints arising from the condition of passivity is evidenced by the fact that several works have been focused on passivity constraints [13-17].

According to the values for *D* and *P* of a given a Mueller matrix **M**, there are the following possibilities, *a*) $P = D = 0$; *b*) $P = D > 0$; c) $D > P$, and d) $P > D$.

Let us first observe that, in the particular case that $P = D = 0$ (i.e. $P_p = 0$), the arbitrary decomposition of can always be performed in such a way that all the parallel components of **M** are orthogonal Mueller matrices (i.e. corresponding to respective retarders), which lack of diattenuation and polarizance, and therefore any **M** of the form

$$\mathbf{M}_0 = m_{00} \begin{pmatrix} 1 & \mathbf{0}^T \\ \mathbf{0} & \mathbf{m} \end{pmatrix} \quad (17)$$

can always be expressed as

$$\mathbf{M} = m_{00} \sum_{i=1}^{r} p_i \mathbf{M}_{Ri}, \quad \sum_{i=1}^{r} p_i = 1, \quad (18)$$

where $\mathbf{M}_{Ri}$ are Mueller matrices of transparent retarders (hence pure and passive, with respective MIC equal to 1). This result shows that in the case of matrices of the form (17), the necessary passivity conditions (3) take the simple form $m_{00} \leq 1$ and are also sufficient. Therefore, in what follows we will consider only the case where the degree of polarization $P_p$ of $\mathbf{M}$ is nonzero.

When $\mathbf{M}$ exhibits a certain amount of diattenuation or polarizance, the demonstration that (3) are sufficient conditions for $\mathbf{M}$ to be passive is more complicated and requires some additional steps, like the introduction of the notion of passive form of a Mueller matrix. Let us first recall that when passivity constraints are not considered (as for instance in relative polarimetry, where $\mathbf{M}$ is measured up to a positive scale factor) it is common to represent by means of $\hat{\mathbf{M}} \equiv \mathbf{M}/m_{00}$ all the equivalence class of Mueller matrices proportional to $\mathbf{M}$. Nevertheless, $\hat{\mathbf{M}}$ only satisfies the necessary passivity conditions (3) in the particular case that $P = D = 0$ (above considered). From (3) it follows that the less restrictive passive representative of $\mathbf{M}$ is given by

$$\tilde{\mathbf{M}} \equiv \frac{1}{1+X} \begin{pmatrix} 1 & \mathbf{D}^T \\ \mathbf{P} & \mathbf{m} \end{pmatrix}, \quad X \equiv \max(D, P). \quad (19)$$

That is, $\mathbf{M}$ satisfies the necessary passivity conditions (3) if and only if $m_{00} \leq 1/(1+X)$, where $X = D$ when $D \geq P$ and $X = P$ when $P \geq D$, so that $\tilde{\mathbf{M}}$ is the passive representative of $\mathbf{M}$ having the maximal value for $m_{00}$ compatible with passivity, $m_{00(\max)} = 1/(1+X)$. Thus, for the sake of conciseness, we will call $\tilde{\mathbf{M}}$ the *passive form* of $\mathbf{M}$. This name will be fully justified when the fact that conditions (3) are not only necessary, but also sufficient for $\mathbf{M}$ to be passive, is demonstrated in Sec. 8.

## 5. PASSIVE PARALLEL DECOMPOSITIONS OF A MUELLER MATRIX

From the concept of a general Mueller matrix $\mathbf{M}$ as an ensemble average of pure Mueller matrices, it follows that $\mathbf{M}$ is passive if there exists at least one way to express $\mathbf{M}$ as a convex combination of passive pure Mueller matrices. Let us consider the passive form $\tilde{\mathbf{M}}$ of a given Mueller matrix $\mathbf{M}$ and its arbitrary decomposition into passive Mueller representatives $\tilde{\mathbf{M}}_{Ji}$ of a set of $r$ pure components, with $r = \text{rank}\left[\mathbf{C}(\tilde{\mathbf{M}})\right]$ (recall that $r$ is the minimum number of pure parallel components of $\mathbf{M}$ and $\tilde{\mathbf{M}}$)

$$\tilde{\mathbf{M}} = \sum_{i=1}^{r} k_i \tilde{\mathbf{M}}_{Ji}, \quad \sum_{i=1}^{r} k_i = 1, \quad (20)$$

Let us denote $\mathbf{D} = \mathbf{X}$ if $D \geq P$ or $\mathbf{P} = \mathbf{X}$ if $P > D$, $\mathbf{D}$ and $\mathbf{P}$ being the diattenuation and polarizance vectors of $\tilde{\mathbf{M}}$ (i.e. of $\mathbf{M}$) and consider Eq. (20) particularized for the element $\tilde{m}_{00}$ and for vector $\mathbf{X}$

$$\frac{1}{1+X} = \sum_{i=1}^{r} k_i \frac{1}{1+X_i},$$

$$\frac{\mathbf{X}}{1+X} = \sum_{i=1}^{r} k_i \frac{\mathbf{X}_i}{1+X_i}, \quad (21)$$

$$k_i = p_i \frac{1+X_i}{1+X}, \quad \sum_{i=1}^{r} p_i = 1,$$

where $\mathbf{X}_i$ are the diattenuation or polarizance vectors of $\tilde{\mathbf{M}}_{Ji}$ depending on if $D \geq P$ or $P \geq D$ respectively. Therefore, by combining these equations, we get

$$\mathbf{X} = \sum_{i=1}^{r} p_i \mathbf{X}_i, \quad \sum_{i=1}^{r} p_i = 1. \quad (22)$$

For the demonstration that conditions (3) are sufficient for $\tilde{\mathbf{M}}$ to be expressed, at least in one form, as a convex combination of passive pure Mueller matrices $\tilde{\mathbf{M}}_{Ji}$, we are interested in identifying the specific decomposition (20) for which the constraints on the passivity are less restrictive, that is, for which $X_i$ take the smaller possible values. Since Eq. (22) represents a sum of vectors $p_i \mathbf{X}_i$, this occurs necessarily when all these vectors are mutually parallel and with the same direction as that of the resultant vector $\left(p_i \mathbf{X}_i \uparrow\uparrow \mathbf{X}\right)$, which in its turn implies that $\mathbf{X}_i \uparrow\uparrow \mathbf{X}$ (recall that $p_i > 0$). This result will be key for the demonstration of the sufficiency of conditions (3) for $\tilde{\mathbf{M}}$ to be passive.

## 6. TRIDIAGONAL FORM OF A MUELLER MATRIX

Given a Mueller matrix $\mathbf{M}$ and an arbitrary pair of orthogonal Mueller matrices $(\mathbf{M}_{RI}, \mathbf{M}_{RO})$, we can consider the *dual-retarder transformation* [28]

$$\begin{aligned}\mathbf{M}' &= \mathbf{M}_{RO} \mathbf{M} \mathbf{M}_{RI} = \begin{pmatrix} 1 & \mathbf{0}^T \\ \mathbf{0} & \mathbf{m}_{RO} \end{pmatrix} m_{00} \begin{pmatrix} 1 & \mathbf{D}^T \\ \mathbf{P} & \mathbf{m} \end{pmatrix} \begin{pmatrix} 1 & \mathbf{0}^T \\ \mathbf{0} & \mathbf{m}_{RI} \end{pmatrix} \\ &= m_{00} \begin{pmatrix} 1 & \mathbf{D}^T \mathbf{m}_{RI} \\ \mathbf{m}_{RO} \mathbf{P} & \mathbf{m}_{RO} \mathbf{m} \mathbf{m}_{RI} \end{pmatrix}.\end{aligned} \quad (23)$$

where $\mathbf{M}_{RI}$ and $\mathbf{M}_{RO}$ represent respective retarders, so that they lack of polarizance-diattenuation and their 3×3 submatrices $\mathbf{m}_{RI}$ and $\mathbf{m}_{RO}$ are proper orthogonal matrices (i.e. $\det \mathbf{m}_{R1} = \det \mathbf{m}_{R2} = +1$). Matrices $\mathbf{M}'$ obtained from $\mathbf{M}$ by means of this kind of transformation are said to be *invariant-equivalent* to $\mathbf{M}$ because $\mathbf{M}'$ and $\mathbf{M}$ share ten invariant properties [28], two of them being $D$ and $P$.

Since the necessary passivity conditions (3) only depend on the absolute values, $D$ and $P$, of the diattenuation and polarizance vectors of $\mathbf{M}$, the expressions for such conditions are preserved under dual-retarder transformations.

In particular, $\mathbf{M}_{RI}$ and $\mathbf{M}_{RO}$ can always be chosen in such a manner that the transformed matrix takes the tridiagonal form [29]

$$\mathbf{M}_t = m_{00} \begin{pmatrix} 1 & D & 0 & 0 \\ P & x_{11} & x_{12} & 0 \\ 0 & x_{21} & x_{22} & x_{23} \\ 0 & 0 & x_{32} & x_{33} \end{pmatrix}. \quad (24)$$

Note that the sings of the transformed elements $x_{01} = D \geq 0$ $x_{10} = P \geq 0$ have been taken positive, which is realizable through the appropriate choice of $\mathbf{M}_{RI}$ and $\mathbf{M}_{RO}$ (the resulting sign of $x_{11}$ being fixed by the said choice). In further sections we will take advantage of this simplified form, which always allows to retrieve $\mathbf{M}$ through the complementary, and reversible (i.e. not involving diattenuation or polarizance effects), dual-retarder transformation $\mathbf{M} = \mathbf{M}_{RO}^T \mathbf{M}_t \mathbf{M}_{RI}^T$.

The passive form $\tilde{\mathbf{M}}_t$ of the tridiagonal Mueller matrix $\tilde{\mathbf{M}}_t$ is given by $\tilde{\mathbf{M}}_t = (1/1 + X) \hat{\mathbf{M}}_t$, with $X \equiv \max(D, P)$.

From the general expressions of the elements of the coherency matrix $\mathbf{C}$ in terms of those of the associated Mueller matrix $\mathbf{M}$ [21], the elements of $\mathbf{C}_t$ (associated with $\tilde{\mathbf{M}}_t$) are given by

$$c_{00} = \frac{m_{00}}{4}(1 + x_{11} + x_{22} + x_{33}),$$

$$c_{01} = c_{10}^* = \frac{m_{00}}{4}[D + P - i(x_{23} - x_{32})],$$

$$c_{02} = c_{20}^* = 0,$$

$$c_{03} = c_{30}^* = -i\frac{m_{00}}{4}(x_{12} - x_{21}),$$

$$c_{11} = \frac{m_{00}}{4}(1 + x_{11} - x_{22} - x_{33}), \quad (25)$$

$$c_{12} = c_{21} = x_{12} + x_{21},$$

$$c_{13} = c_{31}^* = 0,$$

$$c_{22} = \frac{m_{00}}{4}(1 - x_{11} + x_{22} - x_{33}),$$

$$c_{23} = c_{32}^* = \frac{m_{00}}{4}[x_{23} + x_{32} + i(D - P)],$$

$$c_{33} = \frac{m_{00}}{4}(1 - x_{11} - x_{22} + x_{33}).$$

## 7. COHERENCY VECTORS HAVING PARALLEL DIATTENUATION VECTORS

As seen in Sec. 5, the passivity constraints for parallel decompositions features the most relaxed limits when $\mathbf{X}_i \uparrow\uparrow \mathbf{X}$ and the aim of this section is to formulate the expression of a coherency vector $\mathbf{c}$ whose associated diattenuation vector $\mathbf{D}_i$ satisfies the property $\mathbf{D}_i \uparrow\uparrow \mathbf{D}_t$ ($\mathbf{D}_t$ being the diattenuation vector of $\mathbf{M}_t$). Hereafter for the sake of clarity, we will suppose that $D > P$ because, as we will see in Sec. 8, the case $D = P$ does not require further developments (in this case $\mathbf{M}$ can be decomposed into a set of $r - 1$ retarders and a single nonnormal diattenuator whose diattenuation and polarizance vectors have equal magnitudes and are parallel to the diattenuation and polarizance vectors of $\mathbf{M}$), while the case $P > D$ can be treated in fully analogy to the case $D > P$, but considering $\mathbf{M}^T$ (or $\mathbf{M}_t^T$) instead of $\mathbf{M}$ (or $\mathbf{M}_t$).

Given a pure Mueller matrix $\mathbf{M}_J$ its associated pure coherency matrix $\mathbf{C}_J$ can be expressed as $\mathbf{C}_J = \mathbf{c} \otimes \mathbf{c}^\dagger$ in terms of the *coherency vector* $\mathbf{c} = \sqrt{m_{00}}\,\hat{\mathbf{u}}_1$, where $m_{00}$ is the MIC of $\mathbf{M}_J$, $\hat{\mathbf{u}}_1$ is the only unit eigenvector of $\mathbf{C}_J$ with nonzero eigenvalue $\lambda_1$ ($\lambda_1 = \mathrm{tr}\,\mathbf{C}_J = m_{00}$). $\mathbf{c}$ is linked to the covariance vector $\mathbf{h}$ [21,30] that defines the pure covariance matrix $\mathbf{H}_J = \mathbf{h} \otimes \mathbf{h}^\dagger$ by means of $\mathbf{c} = \mathcal{L}\mathbf{h}$, where $\mathcal{L}$ is the unitary matrix defined in (4). In addition, vector $\mathbf{h} \equiv (h_1, h_2, h_3, h_4)^T$ is directly liked to the Jones matrix $\mathbf{T}$ associated with $\mathbf{M}_J$ through the expression

$$\mathbf{T}(\mathbf{M}_J) = \sqrt{2}\begin{pmatrix} h_1 & h_2 \\ h_3 & h_4 \end{pmatrix} \quad (26)$$

For simplicity of further mathematical expressions we will take advantage of the tridiagonal form $\mathbf{M}_t$ of $\mathbf{M}$ and its parallel decomposition into pure Mueller matrices whose diattenuation vectors are parallel to that of $\mathbf{M}_t$. Thus, we are now interested in obtaining the general form of a Jones matrix $\mathbf{T}_\uparrow$ whose corresponding diattenuation vector has the form $\mathbf{D}_\uparrow = (D_\uparrow, 0, 0)^T$, with $D_\uparrow = |\mathbf{D}_\uparrow| > 0$. To do so, let us now recall that the singular value decomposition (6), where the central diattenuating matrix $\mathbf{T}_{DL0} \equiv \mathrm{diag}(p_1, p_2)$ involves diattenuation and polarizance vectors whose only nonzero component is the first one. The unitary matrix $\mathbf{T}_{R1}$ produces the effect of changing the spatial orientation of the diattenuation vector of $\mathbf{T}_{DL0}$ (except for the trivial case in which $\mathbf{T}_{R1}$ coincides with the 2×2 identity matrix $\mathbf{I}_2$). Analogously, $\mathbf{T}_{R2}$ produces the effect of changing the spatial orientation of the polarizance vector of $\mathbf{T}_{DL0}$ (except when $\mathbf{T}_{R2} = \mathbf{I}_2$). Therefore, $\mathbf{T}_\uparrow$ can always be written as $\mathbf{T} = \mathbf{T}_{R2}\,\mathbf{T}_{DL0}$, and by considering the general form of a unitary Jones matrix in terms of three angular parameters $(\alpha, \delta, \Delta)$ [31] we get

$$\mathbf{T}_\uparrow = p_1 \mathbf{T}_R(\alpha, \delta, \Delta)\begin{pmatrix} 1 & 0 \\ 0 & g \end{pmatrix} = $$
$$= p_1 \begin{pmatrix} c_\alpha^2 e^{i\Delta/2} + s_\alpha^2 e^{-i\Delta/2} & igs_{2\alpha}s_{\Delta/2}e^{-i\delta} \\ is_{2\alpha}s_{\Delta/2}e^{i\delta} & g\,s_\alpha^2 e^{i\Delta/2} + g\,c_\alpha^2 e^{-i\Delta/2} \end{pmatrix}, \quad (27)$$

where $g \equiv p_2/p_1$ and the concise notations $s_\varphi \equiv \sin\varphi$ and $c_\varphi \equiv \cos\varphi$ are used. The corresponding covariance and coherency vectors are

$$\mathbf{h}_\uparrow = (p_1/\sqrt{2})\begin{pmatrix} c_\alpha^2 e^{i\Delta/2} + s_\alpha^2 e^{-i\Delta/2} \\ igs_{2\alpha}s_{\Delta/2}e^{-i\delta} \\ is_{2\alpha}s_{\Delta/2}e^{i\delta} \\ g\,s_\alpha^2 e^{i\Delta/2} + g\,c_\alpha^2 e^{-i\Delta/2} \end{pmatrix},$$

$$\mathbf{c}_\uparrow = \mathcal{L}\mathbf{h}_\uparrow = \frac{p_1}{2}\begin{pmatrix} c_\alpha^2\left(e^{i\Delta/2} + g e^{-i\Delta/2}\right) + s_\alpha^2\left(e^{-i\Delta/2} + g e^{-i\Delta/2}\right) \\ c_\alpha^2\left(e^{i\Delta/2} - g e^{-i\Delta/2}\right) + s_\alpha^2\left(e^{-i\Delta/2} - g e^{-i\Delta/2}\right) \\ is_{2\alpha}s_{\Delta/2}\left(e^{i\delta} + g e^{-i\delta}\right) \\ s_{2\alpha}s_{\Delta/2}\left(e^{i\delta} - g e^{-i\delta}\right) \end{pmatrix}. \quad (28)$$

By writing the above expressions of the elements of $\mathbf{c}_\uparrow$ in terms of real and imaginary parts, it follows that $\mathbf{c}_\uparrow$ exhibits the following characteristic structure

$$\mathbf{c}_\uparrow = (q_1 + i k q_2, k q_1 + i q_2, k q_3 + i q_4, -k q_4 + i q_3)^T, \quad (29)$$

where $k$ and $q_i$ ($i = 1, 2, 3, 4$) are real parameters, that is, $\mathbf{c}_\uparrow$ can be written as in Eq. (29) if and only if it has the form shown in Eq. (28). Note that the limiting value $k = 0$ corresponds to the case where $\mathbf{X}_i = \mathbf{0}$, i.e., vector $\mathbf{c}_\uparrow$ is associated to a retarder, which, as shown in [29] (see Sec. 8 of the present paper), can be considered a component of $\mathbf{C}_t$ (with rank $\mathbf{C}_t = 2$) if and only if $P = D$.

Now let us now bring up the generalized arbitrary decomposition (16) formulated in terms of coherency matrices

$$\mathbf{C} = \sum_{i=1}^r k_i \mathbf{C}_{Ji}, \quad \mathbf{C}_{Ji} = \mathbf{c}_i \otimes \mathbf{c}_i^\dagger,$$

$$k_i = \frac{m_{00}}{m_{00i}\sum_{j=1}^r \frac{1}{\lambda_j}\left|(\mathbf{U}^\dagger \mathbf{c}_i)_j\right|^2}, \quad \sum_{i=1}^r k_i = 1. \quad (30)$$

where the vector $\mathbf{c}_i$ generating the corresponding pure component $\mathbf{C}_{Ji} = \mathbf{c}_i \otimes \mathbf{c}_i^\dagger$ of $\mathbf{C}$, necessarily satisfies $\mathbf{c}_i \in \mathrm{range}(\mathbf{C})$, and therefore there always exists a vector $\mathbf{y}_i$ (in general not unique) such that $\mathbf{c}_i = \mathbf{C}\mathbf{y}_i$.

## 8. SUFICIENCE OF THE PASSIVITY CONDITIONS

To perform the demonstration that conditions (3) are sufficient for $\mathbf{M}$ to be a passive Mueller matrix (i.e. $\mathbf{M}$ can be expressed as a convex combination of passive pure Mueller matrices), let us consider separately the cases corresponding to the possible values of $r \equiv \mathrm{rank}\,\mathbf{C}$. It has been proven recently [29] that, for $r = 3, 4$ a respective number $q = 1, 2$ of retarders can be identified as incoherent components of $\mathbf{M}$ and, if $P = D$, then $q = 2, 3$ respectively (note that retarders exhibit zero diattenuation vector

**0**, which can be considered as a limiting case of a diattenuation vector that is parallel to another given diattenuation vector **D**). Furthermore, when $r = 2$ and $P = D$, then **M** can be decomposed as follows in terms of a retarder $\mathbf{M}_R$ (hence passive) and a passive pure Mueller matrix $\tilde{\mathbf{M}}_J$ [29]

$$\tilde{\mathbf{M}} \equiv p\,\mathbf{M}_R + (1-p)\,\tilde{\mathbf{M}}_J . \qquad (31)$$

The case $P = D = 0$, where the necessary and sufficient passivity conditions become trivial $m_{00} > 0$, has already been considered in Sec. 4.

Therefore, the only remaining case to be considered is $r = 2$ with $P \neq D$. For such case, let us take an arbitrary coherency vector $\mathbf{y} \equiv (y_1, y_2, y_3, y_4)^T$ and note that the vector **z** obtained as $\mathbf{z} = \mathbf{C}_t \mathbf{y}$ necessarily belongs to $\mathrm{range}(\mathbf{C}_t)$, that is, for any vector satisfying $\mathbf{z} \in \mathrm{range}(\mathbf{C}_t)$ always exists a vector **y** such that $\mathbf{z} = \mathbf{C}_t \mathbf{y}$ (note that, in general, **y** is not unique). Let us use the expressions (25) to impose that vector $\mathbf{C}_t \mathbf{y}$ has the required form (29) for its associated pure Mueller matrix $_T\tilde{\mathbf{M}}_{J1}$ to have a diattenuation vector of the form $\mathbf{D}_{J1}^T = (D_{J1}, 0, 0)^T$

$$\mathbf{C}_t \mathbf{y} = m_{00} \begin{pmatrix} q_1 + ikq_2 \\ kq_1 + iq_2 \\ kq_3 + iq_4 \\ -kq_4 + iq_3 \end{pmatrix}, \qquad (32)$$

so that, by equating real and imaginary parts of the respective components of **z** vector in both sides of Eq. (32) and by imposing conditions for **z** to have the required form

$$k = \mathrm{Re}(z_2)/\mathrm{Re}(z_1) = \mathrm{Im}(z_1)/\mathrm{Im}(z_2) =$$
$$= \mathrm{Re}(z_3)/\mathrm{Im}(z_4) = -\mathrm{Re}(z_4)/\mathrm{Im}(z_3), \qquad (33)$$

a set of four equations (with $m_{00} \neq 0$) is obtained in terms of the eight variables constituted by the real and imaginary parts $a_i, b_i$ of the respective complex elements $y_i = a_i + ib_i$ of vector **y**. Obviously we are considering $m_{00} \neq 0$, otherwise the overall matrices **M** and $\mathbf{M}_t$ vanish and the problem has no sense. Then, by isolating the variables $(a_1, a_2, a_3, b_4)$ and writing them in terms of the four remaining variables $(b_1, b_2, b_3, a_4)$, we get

$$a_i = \frac{\Delta A_i}{\Delta(D^2-P^2)(x_{23}^2-x_{32}^2)} \quad (i=1,2,3),$$
$$b_4 = \frac{\Delta B_4}{\Delta(D^2-P^2)(x_{23}^2-x_{32}^2)}, \qquad (34)$$

where

$$A_1 \equiv -b_1(D^2-P^2)(D+P)(x_{23}+x_{32})-$$
$$-b_2(D^2-P^2)(x_{23}+x_{32})(1+x_{11}-x_{22}-x_{33})+$$
$$+b_3(x_{12}+x_{21})\left[-D^2 x_{32}+P^2 x_{23}+P(D+P)x_{32}\right],$$

$$A_2 \equiv b_2(D^2-P^2)(D+P)(x_{23}+x_{32})+$$
$$+b_1(D^2-P^2)(x_{23}+x_{32})(1+x_{11}+x_{22}+x_{33})+ \qquad (35)$$
$$+a_4(x_{12}-x_{21})\left[-D^2 x_{32}+P^2 x_{23}+P(D+P)x_{32}\right],$$

$$A_3 \equiv -b_3(D^2-P^2)(D-P)(x_{23}-x_{32})-$$
$$-a_4(D^2-P^2)(x_{23}-x_{32})(1-x_{11}-x_{22}+x_{33})-$$
$$-b_1(x_{12}-x_{21})\left[D^2 x_{32}+P^2 x_{23}+P(D-P)x_{32}\right],$$

$$B_4 \equiv -a_4(D^2-P^2)(D-P)(x_{23}-x_{32})-$$
$$-b_3(D^2-P^2)(x_{23}-x_{32})(1-x_{11}+x_{22}-x_{33})+$$
$$+b_2(x_{12}+x_{21})\left[D^2 x_{32}+P^2 x_{23}+P(D-P)x_{32}\right],$$

$$\Delta \equiv (x_{23}^2-x_{32}^2)\left[\begin{array}{l}(1+k^2)^2(D^2-P^2)+\\ +4(k+k^3)(P-Dx_{11})-4k^2(1-x_{11}^2-x_{12}^2)\end{array}\right].$$

Note that, provided the compatibility of the equations is preserved, arbitrary values can be given for the four free variables $b_1, b_2, b_3, a_4$, leading to respective different solutions, which can adopt simple forms.

The denominator in Eqs. (34) involves $\Delta$, $D^2 - P^2$ and $x_{23}^2 - x_{32}^2$. From the starting hypothesis, $P \neq D$ and therefore $D^2 - P^2 \neq 0$. The particular case $x_{23}^2 = x_{32}^2$ is considered in Appendix B. Observe also that $\Delta$ appears in both the numerator and denominator of the expressions (34), so that, provided $\Delta \neq 0$, it can be simplified. If $\Delta \neq 0$, then it follows that the expression of vector $\mathbf{z} = \mathbf{C}_t \mathbf{y}$ in Eq. (32) results in $\mathbf{z} = \mathbf{0}$, showing that $\Delta \neq 0$ implies that the Mueller matrix associated with **z** is just the zero matrix (which, obviously, is not a valid solution for our purposes). Therefore, the only possibility of finding solutions for vector **z** with $\mathbf{z} \in \mathrm{range}(\mathbf{C})$ and $\mathbf{z} \neq \mathbf{0}$ entails that $\Delta = 0$ (recall that we are considering $P \neq D$ and $x_{23}^2 \neq x_{32}^2$).

Equation $\Delta = 0$ leads to the following four real solutions for parameter $k$

$$k_1 = \frac{1}{y}\left[(-1+\sqrt{x})+\sqrt{-y^2+(1-\sqrt{x})^2}\right],$$
$$k_2 = \frac{1}{y}\left[(-1+\sqrt{x})-\sqrt{-y^2+(1-\sqrt{x})^2}\right],$$
$$k_3 = \frac{1}{y}\left[(-1-\sqrt{x})+\sqrt{-y^2+(1+\sqrt{x})^2}\right], \qquad (36)$$
$$k_4 = \frac{1}{y}\left[(-1-\sqrt{x})-\sqrt{-y^2+(1+\sqrt{x})^2}\right],$$

where

$$x \equiv 1 + \frac{(D^2-P^2)(1-x_{11}^2-x_{12}^2)}{(P-Dx_{11})^2},$$
$$y \equiv \frac{(D^2-P^2)}{P-Dx_{11}}. \qquad (37)$$

To ensure the compatibility of these solutions, it is necessary to solve separately the case where $P = Dx_{11}$ [see the denominators in Eqs. (37)] and also to demonstrate that the radicands in the expressions (36) for the four roots $k_i$ are real. The said required demonstrations are included in Appendix A.

Let us now remember that, in order to complete the demonstration that conditions (3) are sufficient for **M** to be passive, we should analyze the particular case that $x_{23}^2 = x_{32}^2$. Since we are considering the coherency matrix $\mathbf{C}_t$ with $\mathrm{rank}\,\mathbf{C}_t = \mathrm{rank}\,\mathbf{C} = 2$, it follows that all order-3 minors of $\mathbf{C}_t$ are necessarily zero, which in turn entails that $x_{12} = x_{21} = 0$, so that the tridiagonal Mueller matrix $\mathbf{M}_t$ adopts a particularly simple form. Now we proceed similarly to the previous case, but here it results advantageous and simpler to obtain the expressions for the variables $(a_2, a_3, b_1, b_3)$ as linear functions of $(a_1, a_4, b_2, b_4)$. In order to get specific solutions, the subcases (1) $x_{23} = x_{32}$ and (2) $x_{23} = -x_{32}$ are considered separately, and solved in Appendix B.

Once it has been proven the existence of physically realizable solutions for passive decompositions of the tridiagonal form $\mathbf{M}_t$ of a given Mueller matrix $\mathbf{M}$ satisfying rank $\mathbf{C}(\mathbf{M}) = 2$, and $D > P$, this result also applies to $\mathbf{M}$ because the dual retarder transformations (23) do not affect the passivity conditions (3). Furthermore, the sufficiency of such passivity conditions for the case where $P > D$ can be demonstrated through the procedure followed for $\mathbf{M}$, but replacing $\mathbf{M}$ by $\mathbf{M}^T$.

## 9. CONCLUSION

Passivity (non amplification of the intensity of light) is a natural behavior of polarimetric samples that entails certain conditions to be satisfied by Mueller matrices representing material samples. Therefore, a complete mathematical characterization of Mueller matrices requires the identification of a complete minimum set of passivity conditions as well as their rigorous demonstration. While the fact that conditions

$$m_{00}(1+D) \leq 1, \quad m_{00}(1+P) \leq 1, \tag{38}$$

are necessary for a Mueller matrix $\mathbf{M}$ to be passive, the lack of a complete demonstration of their sufficiency has originated certain controversies [8,21].

In the case of pure Mueller matrices, it results obvious that conditions (38) are necessary and sufficient for passivity. Nevertheless, in the case of depolarizing Mueller matrices the sufficiency requires that the fact that a Mueller matrix $\mathbf{M}$ satisfies the inequalities (38) implies that there is at least one way to express $\mathbf{M}$ as a convex composition of passive pure Mueller matrices. This problem has been solved in this work through the procedure indicated below, which additionally has involved new interesting concepts like the *passive form* and the *tridiagonal form* of $\mathbf{M}$ as well as the generalized arbitrary decomposition of $\mathbf{M}$ in terms of passive forms of the Mueller matrices involved.

Given a Mueller matrix $\mathbf{M}$, it can be classified into one of the following types with respect to its diattenuation-polarizance properties, (a) $D = P = 0$; (b) $D = P > 0$, and (c) $D \neq P$.

In [29] it has been proven that any Mueller matrix of type (a) can be considered as a parallel (or incoherent) combination of pure Mueller matrices associated with retarders, in which case the passivity conditions become the trivial single necessary and sufficient passivity condition $m_{00} \leq 1$.

Furthermore, in [29] it has also been proven that any Mueller matrix of type (b) can be decomposed as a convex combination of a set of $r-1$ Mueller matrices of retarders [with $r \equiv$ rank $\mathbf{C}(\mathbf{M})$] and one pure Mueller matrix that accumulates all the diattenuation and polarizance charge of the components, so that the sufficiency of the necessary passivity conditions (38) is directly satisfied. The remaining case (c) is thus reduced to Mueller matrices satisfying $r = 2$ and $D \neq P$, for which the sufficiency of the necessary passivity inequalities (38) has been proven in this work for the first time.

Therefore, the complete characterization of passive Mueller matrices is attained by means of two sets of inequalities, namely the four covariance conditions provided by the nonnegativity of the eigenvalues of the coherency matrix $\mathbf{C}$ associated with a given Mueller matrix $\mathbf{M}$, and the pair of passivity conditions (38).

## Appendix A

Let us first analyze the particular case where the quantity $P - Dx_{11}$ appearing in the denominators of the expressions that define $y$ as well as the second term of $x$ in Eq. (37) is zero. When $P = Dx_{11}$ then $\Delta$ takes the particular form

$$\Delta \equiv (x_{23}^2 - x_{32}^2)\left[(1+k^2)^2(D^2 - P^2) - 4k^2(1 - x_{11}^2 - x_{12}^2)\right], \tag{A1}$$

and the solutions obtained for the four roots $k_i$ of the equation $\Delta = 0$ are now the following

$$k_1 = \frac{\sqrt{1 - x_{11}^2 - x_{12}^2} + \sqrt{1 - D^2 + P^2 - x_{11}^2 - x_{12}^2}}{\sqrt{D^2 - P^2}},$$

$$k_2 = \frac{\sqrt{1 - x_{11}^2 - x_{12}^2} - \sqrt{1 - D^2 + P^2 - x_{11}^2 - x_{12}^2}}{\sqrt{D^2 - P^2}},$$

$$k_3 = \frac{-\sqrt{1 - x_{11}^2 - x_{12}^2} + \sqrt{1 - D^2 + P^2 - x_{11}^2 - x_{12}^2}}{\sqrt{D^2 - P^2}},$$

$$k_4 = \frac{-\sqrt{1 - x_{11}^2 - x_{12}^2} - \sqrt{1 - D^2 + P^2 - x_{11}^2 - x_{12}^2}}{\sqrt{D^2 - P^2}}.$$

(A2)

where 1) the radicand in the denominator is positive $D^2 - P^2 > 0$ because of the starting hypothesis $D > P$ (recall that the case $D = P$ has been previously studied separately in Ref. [29]); 2) the radicand $1 - x_{11}^2 - x_{12}^2$ is nonnegative because any Mueller matrix $\mathbf{M}$ with elements $m_{ij}$ $(i, j = 0,1,2,3)$ satisfies the property $m_{00}^2 \geq m_{j1}^2 + m_{j2}^2 + m_{j3}^2$ $(j = 0,1,2,3)$ [4] (this property can be demonstrated from the fact that the Stokes vectors obtained as $\mathbf{M}\mathbf{s}_{i\pm}$, $\mathbf{s}_{i\pm}$ being the canonical Stokes vectors $\mathbf{s}_{1\pm} \equiv (1, \pm 1, 0, 0)^T$, $\mathbf{s}_{2\pm} \equiv (1, 0, \pm 1, 0)^T$, $\mathbf{s}_{3\pm} \equiv (1, 0, 0, \pm 1)^T$, can be combined into the Stokes vectors $\mathbf{s}_i = \mathbf{s}_{i+} + \mathbf{s}_{i-}$ [19,32]), and 3) the nonnegativity of the radicand $1 + P^2 - D^2 - x_{11}^2 - x_{12}^2$ is demonstrated by considering the Stokes vector $\mathbf{s}$ obtained through the Mueller-Stokes transformation $\mathbf{s} = \mathbf{M}_t^T (1, -1, 0, 0)^T = (1 - P, D - x_{11}, -x_{12}, 0)^T$, which necessarily satisfies the Stokes vectors condition

$$0 \leq s_0^2 - s_1^2 - s_2^2 - s_3^2 =$$
$$= 1 - D^2 + P^2 + 2Dx_{11} - 2P - x_{11}^2 - x_{12}^2 = \tag{A3}$$
$$= 1 + P^2 - D^2 - x_{11}^2 - x_{12}^2,$$

where $s_i$ $(i = 0,1,2,3)$ are the components of $\mathbf{s}$.

Once the case $P = Dx_{11}$ has been solved, hereafter we will assume that condition $P \neq Dx_{11}$ is satisfied by $\mathbf{M}_t$ and we will inspect the compatibility of the expressions (36) for the roots $k_i$ of the equation $\Delta = 0$, where the radicands $x$, $\left(1 + \sqrt{x}\right) - y^2$ and $\left(1 - \sqrt{x}\right) - y^2$ should be nonnegative in order to get the desired real solutions. From the definition of parameter $x$ in Eq. (37) we see that 1) $D^2 - P^2 > 0$ because of the starting hypothesis $D > P$, and 2) the inequality $1 - x_{11}^2 - x_{12}^2 \geq 0$ has been demonstrated above. Therefore, the required condition $x \geq 0$ is always satisfied.

Concerning the other radicands, observe that $1 + \sqrt{x} \geq 1 - \sqrt{x}$, so that it is enough to show the nonnegativity of $\left(1 - \sqrt{x}\right) - y^2$. Despite the fact that $y$ appears squared, it is worth to distinguish the cases (a) $y = 0$, (b) $y < 0$, and (c) $y > 0$.

Case (a) should be discarded because it is not compatible with the hypothesis $D > P$.

In case (b), condition $\left(1 - \sqrt{x}\right)^2 - y^2 \geq 0$, with $y < 0$ is entirely equivalent to $x \geq (y - 1)^2$. Then, by writing variables $x$ and $y$ in terms of the elements of $\mathbf{M}_t$ and after some mathematical manipulations, the above condition adopts the form

$$\frac{(D^2 - P^2)\left[-(1+P)^2 + (D+x_{11})^2 + x_{12}^2\right]}{(P - D x_{11})^2} \leq 0. \quad \textbf{(A4)}$$

Since $D^2 - P^2 > 0$ (because of the starting hypothesis $D > P$) and the denominator is a squared quantity, the inequality (A4) can be expressed in the following simplified manner (recall that we are considering $P \neq D x_{11}$)

$$(1+P)^2 - (D+x_{11})^2 - x_{12}^2 \geq 0 \quad \textbf{(A5)}$$

which is necessarily satisfied because the Stokes vector obtained as $\mathbf{M}_t^T (1,1,0,0)^T = (1+P, D+x_{11}, x_{12}, 0)^T$ satisfies the Stokes vectors condition $s_0^2 - s_1^2 - s_2^2 - s_3^2 \geq 0$ ($s_i$ being the components of the Stokes vector considered).

In case (c), condition $\left(1 - \sqrt{x}\right)^2 - y^2 \geq 0$, with $y > 0$ is entirely equivalent to $x \geq (1+y)^2$. Then, by writing variables $x$ and $y$ in terms of the elements of $\mathbf{M}_t$ and after some mathematical manipulations, the above condition adopts the form

$$\frac{(D^2 - P^2)\left[-(1-P)^2 + (D-x_{11})^2 + x_{12}^2\right]}{(P - D x_{11})^2} \leq 0. \quad \textbf{(A6)}$$

Since $D^2 - P^2 > 0$ (because of the hypothesis $D > P$) and the denominator is a squared quantity (with $P \neq D x_{11}$), the inequality (A6) can be expressed in the following simplified manner

$$(1-P)^2 - (D-x_{11})^2 - x_{12}^2 \geq 0 \quad \textbf{(A7)}$$

which is necessarily satisfied because the Stokes vector obtained as $\mathbf{M}_t^T (1,-1,0,0)^T = (1-P, D-x_{11}, x_{12}, 0)^T$ satisfies the Stokes vectors condition $s_0^2 - s_1^2 - s_2^2 - s_3^2 \geq 0$.

**Appendix B**

(1) When $x_{23} = x_{32} \neq 0$ (recall that this condition implies that $x_{12} = x_{21} = 0$), the expressions for the isolated variables $(a_2, a_3, b_1, b_3)$ are the following

$$a_2 = -\frac{\Delta' A_2'}{\Delta'(D+P)},$$

$$a_3 = \frac{\Delta' A_3'}{\Delta'(D+P)(1-x_{11}+x_{22}-x_{33})},$$

$$b_1 = -\frac{\Delta' B_1'}{\Delta'(D+P)(1-x_{11}+x_{22}-x_{33})}, \quad \textbf{(B1)}$$

$$b_3 = -\frac{\Delta' B_3'}{\Delta'(1-x_{11}+x_{22}-x_{33})},$$

where

$$A_2' \equiv a_1(1+x_{11}+x_{22}+x_{33}),$$
$$A_3' \equiv -2a_4(D+P)x_{23} + b_4\left(1+x_{11}^2 - x_{22}^2 - x_{33}^2 - 2P^2 - 2x_{23}^2\right),$$
$$B_1' \equiv b_2\left\{D^2 + P^2 + 2\left[x_{11}(-x_{11}+x_{22}) + x_{23}^2 - x_{33}(1-x_{33})\right]\right\},$$
$$B_3' \equiv a_4(D-P) + 2b_4 x_{23}, \quad \textbf{(B2)}$$

$$\Delta' \equiv (D+P)(1-x_{11}+x_{22}-x_{33})\begin{bmatrix}(1+k^2)^2(D^2-P^2) - \\ -4k^2(1-x_{11}^2) + \\ +4(k+k^3)(P-Dx_{11})\end{bmatrix}$$

Recall that, provided the compatibility of the equations is preserved, arbitrary values variables can be given for the four free variables $a_1, b_2, a_4, b_4$, so that the mathematical expressions can adopt simple forms.

The quantity $1 - x_{11} + x_{22} - x_{33}$ in the denominators of $a_3$, $b_1$ and $b_3$ is never zero because otherwise 1) one of the order-3 minors of $\mathbf{C}_t$ is nonzero, which is incompatible with the hypothesis that rank $\mathbf{C}_t = 2$, or 2) $P = D$, against the hypothesis $D > P$. Moreover, $\Delta'$ appears in both the numerator and denominator of the expressions (34), so that, provided $\Delta' \neq 0$, it can be simplified. If $\Delta' \neq 0$, then it follows that the expression of vector $\mathbf{z} = \mathbf{C}_t \mathbf{y}$ in Eq. (32) results in $\mathbf{z} = \mathbf{0}$, showing that $\Delta' \neq 0$ implies that the Mueller matrix associated with $\mathbf{z}$ is just the zero matrix. Therefore, the only possibility of finding solutions for vector $\mathbf{z}$ with $\mathbf{z} \in \text{range}(\mathbf{C})$ and $\mathbf{z} \neq \mathbf{0}$ entails that $\Delta' = 0$.

Equation $\Delta' = 0$ leads to the following four real solutions for parameter $k$

$$k_1 = \frac{1 + x_{11} + \sqrt{(1+x_{11})^2 - (D+P)^2}}{D+P},$$

$$k_2 = \frac{1 + x_{11} - \sqrt{(1+x_{11})^2 - (D+P)^2}}{D+P}, \quad \textbf{(B3)}$$

$$k_3 = \frac{-1 + x_{11} + \sqrt{(1-x_{11})^2 - (D-P)^2}}{D+P},$$

$$k_4 = \frac{-1 + x_{11} - \sqrt{(1-x_{11})^2 - (D-P)^2}}{D+P}.$$

The radicands in the above solutions for $k$ are nonnegative because of the following property satisfied by any Mueller matrix $\mathbf{M}$ [10]

$$(m_{00} + m_{11})^2 \geq (m_{01}+m_{10})^2 + (m_{22}-m_{33})^2 + (m_{23}+m_{32})^2,$$
$$(m_{00} - m_{11})^2 \geq (m_{01}-m_{10})^2 + (m_{22}+m_{33})^2 + (m_{23}-m_{32})^2, \quad \textbf{(B4)}$$

which, when applied to $\mathbf{M}_t$ (with $m_{00} > 0$), implies that

$$(1+x_{11})^2 \geq (D+P)^2, \quad (1-x_{11})^2 \geq (D-P)^2. \quad \textbf{(B5)}$$

(2) When $x_{23} = -x_{32} \neq 0$ (recall that this condition implies that $x_{12} = x_{21} = 0$), the expressions for the isolated variables $(a_2, a_3, b_1, b_3)$ are the following

$$a_2 = -\frac{\Delta' A_2''}{\Delta'(D+P)},$$

$$a_3 = \frac{\Delta' A_3''}{\Delta'(D+P)(1-x_{11}+x_{22}-x_{33})},$$

$$b_1 = \frac{\Delta' B_1''}{\Delta'(D+P)(1-x_{11}+x_{22}-x_{33})}, \quad \textbf{(B6)}$$

$$b_3 = -\frac{\Delta' B_3''}{\Delta'(1-x_{11}+x_{22}-x_{33})},$$

where

$$A_2'' \equiv 2b_2 x_{23} + a_1(1 + x_{11} + x_{22} + x_{33}),$$
$$A_3'' \equiv b_4(1 + x_{11}^2 - x_{22}^2 - x_{33}^2 - 2P^2 - 2x_{23}^2),$$
$$B_1'' \equiv -2a_1 x_{23}(1 - x_{11} + x_{22} - x_{33}) - $$
$$-b_2\{D^2 + P^2 + 2[x_{11}(-x_{11} + x_{22}) + x_{23}^2 - x_{33}(1 - x_{33})]\}$$
$$B_3'' \equiv a_4(D - P). \tag{B7}$$

As in the previous cases, and provided the compatibility of the equations is preserved, arbitrary values variables can be given for the four free variables $a_1, b_2, a_4, b_4$, leading to simple mathematical expressions.

Through the same arguments that as for case (1), equation $\Delta' = 0$ leads to the four real solutions (B3) for parameter $k$.

(3) When $x_{23} = x_{32} = 0$, with $x_{12} + x_{21} \neq 0$, the equations take simple forms when the isolated variables are $(a_1, a_2, b_3, a_4)$, with corresponding expressions

$$a_1 = \frac{\Delta''' A_1'''}{\Delta'''(x_{12} - x_{21})},$$
$$a_2 = -\frac{\Delta''' A_2'''}{\Delta'''(x_{12} - x_{21})(D + P)},$$
$$b_3 = \frac{\Delta''' B_3'''}{\Delta'''(x_{12} + x_{21})}, \tag{B8}$$
$$a_4 = -\frac{\Delta''' A_4'''}{\Delta'''(x_{12} + x_{21})(D - P)},$$

where

$$A_1''' \equiv a_3(D - P) - b_4(1 - x_{11} - x_{22} + x_{33}),$$
$$A_2''' \equiv a_3(D - P)(1 + x_{11} + x_{22} + x_{33})$$
$$+ b_4[(x_{12} - x_{21})^2 + (x_{11} + x_{22})^2 - (1 + x_{33})^2],$$
$$B_3''' \equiv -b_1(D + P) - b_2(1 + x_{11} - x_{22} - x_{33}),$$
$$A_4''' \equiv -b_1(D + P)(1 - x_{11} + x_{22} - x_{33})$$
$$+ b_2[(x_{12} + x_{21})^2 + (x_{11} - x_{22})^2 - (1 - x_{33})^2], \tag{B9}$$

$$\Delta''' \equiv \begin{bmatrix} (D-P)(x_{12}+x_{21})k^2 + \\ +2[x_{21}(1-x_{11}) - x_{12}(x_{22}-x_{33})]k - \\ -(D-P)(x_{12}-x_{21}) \end{bmatrix}$$
$$\begin{bmatrix} (D+P)(x_{12}+x_{21})k^2 + \\ +2[x_{21}(1+x_{11}) + x_{12}(x_{22}+x_{33})]k - \\ -(D+P)(x_{12}+x_{21}) \end{bmatrix}$$

Again, provided the compatibility of the equations is preserved, arbitrary values variables can be given for the four free variables $b_1, b_2, a_3, b_4$, so that the mathematical expressions can adopt simple forms.

As in the cases analyzed previously, when $\Delta''' \neq 0$ then necessarily $\mathbf{z} = \mathbf{0}$, whose associated Mueller matrix is the zero matrix. Therefore, the only possibility of finding solutions for vector $\mathbf{z}$ with $\mathbf{z} \in \mathrm{im}(\mathbf{C})$ and $\mathbf{z} \neq \mathbf{0}$ corresponds to $\Delta''' = 0$, which leads to the following four real solutions for parameter $k$

$$k_1 = -\frac{x_{21}(1 + x_{11}) + x_{12}(x_{22} + x_{33}) - \sqrt{R_1}}{(x_{12} - x_{21})(D + P)},$$
$$k_2 = -\frac{x_{21}(1 + x_{11}) + x_{12}(x_{22} + x_{33}) + \sqrt{R_1}}{(x_{12} - x_{21})(D + P)},$$
$$k_3 = -\frac{x_{21}(1 - x_{11}) - x_{12}(x_{22} - x_{33}) - \sqrt{R_2}}{(x_{12} + x_{21})(D - P)}, \tag{B10}$$
$$k_4 = -\frac{x_{21}(1 - x_{11}) - x_{12}(x_{22} - x_{33}) + \sqrt{R_2}}{(x_{12} + x_{21})(D - P)},$$

where

$$R_1 = (x_{12}^2 - x_{21}^2)(D + P)^2 +$$
$$+ [x_{12}(x_{22} + x_{33}) + x_{21}(1 + x_{11})]^2,$$
$$R_2 = (x_{12}^2 - x_{21}^2)(D - P)^2 +$$
$$+ [x_{12}(x_{22} - x_{33}) - x_{21}(1 - x_{11}) +]^2. \tag{B10}$$

The nonnegativity of the radicands $R_1$ and $R_2$ in the above expressions (and hence the existence of real solutions for $k$) can be demonstrated form the fact that all order-2 minors of $\mathbf{C}_t$ are always nonnegative (recall that $\mathrm{rank}\,\mathbf{C}_t = 2$) and from the Mueller matrices property (B4).

The particular cases where the denominators of the solutions for parameter $k$ in (B10) are zero are analyzed below

(4) With respect to the denominators in (B10), let us note that when $D = P$ (including the limiting values $D = P = 0$), then, as demonstrated in [29], it is always possible to find a retarder as a pure component of $\mathbf{M}_t$ (with $\mathrm{rank}\,\mathbf{C}_t = 2$), so that the sufficiency of the passivity conditions (3) becomes evident. Moreover, when $x_{23} = x_{32} = 0$, with $x_{12} + x_{21} = 0$ and/or $x_{12} - x_{21} = 0$, then the fact that all order-3 minors of $\mathbf{C}_t$ are zero implies that necessarily the equality $x_{12} = x_{21} = 0$., and the equations take simple forms when the isolated variables are $(a_1, b_1, a_4, b_4)$, with corresponding expressions

$$a_1 = -\frac{\Delta^{IV} a_2(D + P)}{\Delta^{IV}(1 + x_{11} + x_{22} + x_{33})},$$
$$b_1 = -\frac{\Delta^{IV} b_2(1 + x_{11} - x_{22} - x_{33})}{\Delta^{IV}(D + P)},$$
$$a_4 = -\frac{\Delta^{IV} b_3(1 - x_{11} + x_{22} - x_{33})}{\Delta^{IV}(D - P)},$$
$$b_4 = \frac{\Delta^{IV} a_3(D - P)}{\Delta^{IV}(1 - x_{11} - x_{22} + x_{33})}, \tag{B11}$$

$$\Delta^{IV} \equiv [D - P + k(1 - x_{11} - x_{22} + x_{33})]$$
$$[D + P - k(1 + x_{11} + x_{22} + x_{33})]$$
$$[k(D + P) - (1 + x_{11} + x_{22} + x_{33})]$$
$$[k(D - P) + (1 - x_{11} - x_{22} + x_{33})].$$

As in the previous cases, and provided the compatibility of the equations is preserved, arbitrary values variables can be given for the four free variables $a_2, b_2, a_3, b_3$, leading to simple mathematical expressions.

Again, when $\Delta^{IV} \neq 0$ then necessarily $\mathbf{z} = \mathbf{0}$, so that the only possibility of finding solutions for vector $\mathbf{z}$ with $\mathbf{z} \in \mathrm{im}(\mathbf{C})$ and $\mathbf{z} \neq \mathbf{0}$ corresponds to $\Delta^{IV} = 0$, which leads to the following four real solutions for parameter $k$

$$k_1 = \frac{D+P}{1+x_{11}+x_{22}+x_{33}},$$

$$k_2 = -\frac{D-P}{1-x_{11}-x_{22}+x_{33}},$$

$$k_3 = \frac{1+x_{11}+x_{22}+x_{33}}{D+P},$$

$$k_4 = -\frac{1-x_{11}-x_{22}+x_{33}}{D-P}.$$

(B12)

When any of the denominators of the above solutions are zero, then the fact that all order-3 minors of $\mathbf{C}_t$ are zero implies that necessarily $D = P$, against our starting hypothesis. Observe that, as indicated above, when $D = P$ it is always possible to find a retarding parallel component of $\mathbf{M}_t$ (with rank $\mathbf{C}_t > 1$) [29].

## References

1. J. J. Gil, E. Bernabéu, "A depolarization criterion in Mueller matrices," Opt. Acta **32**, 259-261 (1985).
2. K. Kim, L. Mandel, E. Wolf "Relationship between Jones and Mueller matrices for random media," J. Opt. Soc. Am. A **4**, 433-437 (1987).
3. S.R. Cloude, "Group theory and polarization algebra", Optik **75**, 26-36 (1986).
4. P. M. Arnal, *Modelo matricial para el estudio de fenómenos de polarización de la luz*. PhD Thesis. University of Zaragoza. 1990.
5. T. Opatrný, J. Peřina, "Non-image-forming polarization optical devices and Lorentz transformations - an analogy," Phys. Lett. A **181**, 199-202 (1993).
6. J. J. Gil, "Characteristic properties of Mueller matrices," J. Opt. Soc. Am. A **17**, 328-334 (2000).
7. R. Ossikovski, "Canonical forms of depolarizing Mueller matrices,", J. Opt. Soc. Am. A **27**, 123-130 (2010).
8. J. J. Gil, I. San José, R. Ossikovski, "Serial-parallel decompositions of Mueller matrices," J. Opt. Soc. Am. A. **30**, 32-50 (2013).
9. J. M. Correas, P. A. Melero, J. J. Gil, "Decomposition of Mueller matrices into pure optical media," Monog. Sem. Mat. G. Galdeano **27**, 23-240 (2003). Available from <http://www.unizar.es/galdeano/actas_pau/PDF/233.pdf>
10. J. J. Gil, "Polarimetric characterization of light and media," Eur. Phys. J. Appl. Phys. **40**, 1-47 (2007).
11. J. J. Gil and I. San José, "Polarimetric subtraction of Mueller matrices," J. Opt. Soc. Am. A **30**, 1078-1088 (2013).
12. Z-F Xing, "On the deterministic and non-deterministic Mueller matrix," J. Mod Opt. **39**, 461-484 (1992).
13. R. Barakat, "Conditions for the physical realizability of polarization matrices characterizing passive systems," J. Mod. Opt. **34**, 1535-1544 (1987).
14. C. Brosseau, R. Barakat, "Jones and Mueller polarization matrices for random media," Opt. Commun. **84**, 127-132 (1991).
15. A. B. Kostinski, R. C. Givens, "On the gain of a passive linear depolarizing system," J. Mod. Opt. **39**, 1947-1952 (1992).
16. V. Devlaminck, P. Terrier, "Non-singular Mueller matrices characterizing passive systems," Optik **121**, 1994-1997 (2010).
17. J. J. Gil, "Transmittance constraints in serial decompositions of Mueller matrices. The arrow form of a Mueller matrix," J. Opt. Soc. Am. A **30**, 701-707 (2013).
18. J. J. Gil, "Components of purity of a Mueller matrix," J. Opt. Soc. Am. A **28**, 1578-1585 (2011).
19. J. J. Gil, *Determination of polarization parameters in matricial representation. Theoretical contribution and development of an automatic measurement device*. PhD thesis, University of Zaragoza, 1983. Available at http://zaguan.unizar.es/record/10680/files/TESIS-2013-057.pdf
20. R. Ossikovski, J. J. Gil, I. San José, "Poincaré sphere mapping of Mueller matrices," J. Opt. Soc. Am. A. **30**, 2291-2305 (2013).
21. J. J. Gil, R. Ossikovski, *Polarized Light and the Mueller Matrix Approach*. (CRC Press 2016).
22. S. R. Cloude, *Polarisation: Applications in Remote Sensing*, (Oxford University Press, 2009).
23. J. J. Gil, I. San José, "Arbitrary decomposition of Mueller matrices," to be published.
24. A. Van Eeckhout, A. Lizana, E. Garcia-Caurel, J. J. Gil, R. Ossikovski, J. Campos, "Synthesis and characterization of depolarizing samples based on the indices of polarimetric purity," Opt. Lett. **42**, 4155-4158 (2017).
25. M. Foldyna, E. Garcia-Caurel, R. Ossikovski, A. De Martino, and J. J. Gil, "Retrieval of a non-depolarizing component of experimentally determined depolarizing Mueller matrices," Opt. Express **17**, 12794-12806 (2009).
26. R. Ossikovski, E. Garcia-Caurel, M. Foldyna, J. J. Gil, "Application of the arbitrary decomposition to finite spot size Mueller matrix measurements," Appl. Opt. **53**, 6030-6036 (2014).
27. J. J. Gil, "Parallel decompositions of Mueller matrices and polarimetric subtraction," EPJ Web of Conferences **5**, 04002-3 (2010).
28. J. J. Gil, "Invariant quantities of a Mueller matrix under rotation and retarder transformations," J. Opt. Soc. Am. A **33**, 52-58 (2016).
29. I.San José, J. J. Gil, "Retarding parallel components of a Mueller matrix," to be published
30. J. J. Gil, "Review on Mueller matrix algebra for the analysis of polarimetric measurements," J. Appl. Remote Sens. **8**, 081599-37 (2014).
31. J. J. Gil and E. Bernabéu, "Obtainment of the polarizing and retardation parameters of a nondepolarizing optical system from its Mueller matrix," Optik **76**, 67-71 (1987).
32. J. J. Gil, E. Bernabéu, "Depolarization and polarization indices of an optical system," Opt. Acta **33**, 185-189 (1986).